\documentclass{emulateapj} 
\usepackage{graphics} 
\usepackage{graphicx,amssymb, txfonts,pstricks}

\def\br{Br$\gamma$}

\def\h2{H$_2$}
\def\p1{Paper~I}
\def\kl{K$_{\rm l}$}

\shorttitle{Intemediate age stars and low-$\sigma_*$ ring in Mrk\,1066}
\shortauthors{Rogemar A. Riffel et al.}

\begin{document}


\title{Intermediate age stars as origin of the low velocity dispersion nuclear ring in Mrk\,1066}


\author{Rogemar A. Riffel$^{1,2}$, Thaisa Storchi-Bergmann$^{2}$, Rog\'erio Riffel$^{2}$ and Miriani G. Pastoriza$^{2}$
}
\affil{$^{1}$Universidade Federal de Santa Maria, Departamento de F\'\i sica, Centro de Ci\^encias Naturais e Exatas, 
97105-900, Santa Maria, RS, Brazil \\ 
$^{2}$Universidade Federal do Rio Grande do Sul, IF, CP 15051, 91501-970, Porto
Alegre, RS, Brazil} \email{rogemar@smail.ufsm.br}

%\and

%\author{Peter J. McGregor
%} \affil{Research School of Astronomy and
%Astrophysics, Australian National University, Cotter Road, Weston Creek,
%ACT 2611, Australia}

%\and

%\author{Tracy Beck} \affil{Gemini Observatory and Space Telescope Science Institute, Baltimore, MD}


%\altaffiltext{1(Visiting Astronomer, Cerro Tololo Inter-American Observatory. 
%CTIO is operated by AURA, Inc.\ under contract to the National Science %Foundation.} 
%\altaffiltext{2(Society of Fellows, Harvard University.} 
%\altaffiltext{3(present address: Center for Astrophysics, 
%    60 Garden Street, Cambridge, MA 02138}
%\altaffiltext{4(Visiting Programmer, Space Telescope Science Institute} 
%\altaffiltext{5(Patron, Alonso's Bar and Grill}


\begin{abstract} 

We report the first two-dimensional stellar population synthesis in the near-infrared of the nuclear region of an active galaxy, namely Mrk\,1066.
 We have used integral field spectroscopy with adaptative optics at the Gemini North Telescope to map the to map 
the age distribution of the stellar population in the inner 300\,pc  at a spatial resolution of 35\,pc. An old stellar population component (age $\gtrsim$5\,Gyr) is dominant within the inner $\approx$\,160\,pc, %contributing with 50\% of the flux at 2.12\,$\mu$m, 
which we attribute to the galaxy bulge. Beyond this region, up to the borders of the observation field ($\sim$\,300\,pc), intermediate age components (0.3--0.7\,Gyr) dominate. 
We find a spatial correlation between this intermediate age component and a partial ring of low stellar velocity dispersions ($\sigma_*$).
 Low-$\sigma_*$ nuclear rings have been observed in other active galaxies and our result for Mrk\,1066 suggests that they 
are formed by intermediate age stars. This age is consistent with an origin for the low-$\sigma_*$  rings in a  past event which 
triggered an inflow of gas  
and formed stars which still keep the colder kinematics (as compared to that of the bulge) of the gas from which they have formed.
At the nucleus proper we detect, in addition, two unresolved components: a compact infrared source, consistent with an origin in hot dust with mass $\approx1.9\times10^{-2}$\,M$_\odot$, and a blue featureless power-law continuum, which contributes with only $\approx$15\% of the flux at  2.12\,$\mu$m.
 
\end{abstract}





\section{Introduction}

Optical spectroscopy on scales of hundreds of parsecs around the nucleus of Seyfert galaxies have shown that in $\approx$\,40\,\% of
 them the active galactic nucleus (AGN) and young stars co-exist \citep[e.g.][]{sb00,sb01,gd01,cid04,asari07,dors08}, %asari07 
providing support to the so-called AGN-Starburst  connection \citep[e.g.][]{norman88,terlevich90,heckman97,heckman04,n7582}.  
In particular, these authors have pointed out that the main difference between the stellar population (hereafter SP) of active 
and non-active galaxies is an excess of mainly intermediate age stars in the former.
%Thus, the study of the stellar populations (SPs) and its contribution to the circumnuclear continuum of active galaxies may be 
%a fundamental key in the understanding the nature of their central engine.

A similar result has been found in recent SP studies in the near-infrared (hereafter near-IR), using the technique 
of spectral synthesis \citep{rogerio09,rogerio07}.  Using integrated spectra of the central few hundreds of parsecs in
 a sample of 24 Seyfert galaxies, these authors have shown that the continuum  is dominated by the contribution of 
intermediate-age stellar population components (SPCs).  In addition, they found that the near-IR nuclear spectra of about
 50\,\% of the Seyfert~1  and $\sim$20\,\% of the Seyfert~2  galaxies show emission from hot dust 
\citep{rogerio09,n4151,n7582,ardila06,ardila05}.

Using a somewhat distinct technique, \citet{davies07} obtained near-IR integral field spectroscopy to investigate the circumnuclear star formation in 9 nearby Seyfert galaxies at spatial resolutions of tens of parsecs. They have modelled the \br\ equivalent width, supernova rate and mass-to-light ratio to quantify the star formation history in the center of these galaxies using their code  {\sc stars}.  They found that the ages of the stars which contribute most to the near-IR continuum lie in the range 10--30\,Myr, but point out that  these ages should be considered only as ``characteristic", as they have not performed a proper spectral synthesis, arguing that there may be simultaneously two or more SPs that are not coeval \citep{davies07,davies06}. 

In this paper we present, for the first time, two-dimensional (hereafter 2D) SP synthesis in the near-IR for the inner hundreds of pc of an active galaxy -- namely Mrk\,1066 -- using integral field spectroscopy with adaptive optics, which  allowed us to derive the contribution of distinct SPCs to the near-IR spectra and map their spatial distributions. This paper is organized as follows. In Sec.~\ref{data} we describe the observations, data reduction procedures and the spectral synthesis method; in Sec.~\ref{results} we present our results, which are discussed in Sec.~\ref{discussion}. 
The conclusions are presented in Sec.~\ref{disc}.

\section{The data and synthesis code} \label{data}
\subsection{Mrk\,1066}

Mrk 1066 is an SB0 galaxy harboring a Seyfert 2 nucleus and located at a distance
of 48.6 Mpc, for which 1\arcsec corresponds to 235 pc at the galaxy. Previous studies of the central
region of Mrk 1066 have shown that line emission from high-excitation gas is dominated by
gas in a bi-conical outflow oriented along position angle
 PA=135/315$^\circ$, which seems to be associated to the radio jet, while the low-excitation 
gas is more restricted to the plane of the galaxy \citep{bower95,nagar99,knop01,paper1,paper2}.
We selected this galaxy for the present study because it shows a signature of the presence of young/intermediate age stellar
 population in its stellar kinematics in the inner few hundred parsecs as illustrated in the stellar velocity dispersion
 ($\sigma_*$) map shown in the top-left panel of Fig.~\ref{pop}. This map was obtained from the fit
 of the CO band heads around 2.3\,$\mu$m by a combination of stellar spectral templates
 from \citet{winge09}  using the penalized Pixel-Fitting (pPXF) method of \citet{cappellari04}. It shows a partial ring of low $\sigma_*$, 
a signature of the presence of stars which still have the kinematics of the cold gas from which they have 
formed\footnote{A detailed discussion of the stellar kinematics is presented in \citet{paper2}.}. 
Our first goal with the spectral synthesis is to derive the age of this population.  
Additionally, high spatial resolution spectroscopy has shown that the  near-IR nuclear continuum is consistent with a
 large contribution from hot dust emission \citep{paper1}, not detected in lower resolution observations \citep{rogerio09}. 
Our second goal is to quantify this contribution with spectral synthesis. %Previous studies 
%of the circumnuclear region of Mrk\,1066  show that the high- and low-ionization gas present distinct kinematics and origin, with the former dominated by emission 
%in an outflowing gas associated with the radio jet and the latter by emission from the plane of the galaxy \citep{paper1,paper2,knop01,bower95}. 
%In \citet{paper1} we present a detailed discussion of the near-IR emission-line flux distributions and gaseous excitation, while in \citet{paper2} we 
%present the stellar and gaseous kinematics of the circumnuclear region of Mrk\,1066.

%\begin{figure}
%\centering
%\includegraphics[scale=0.5]{figs/sig.ps}
%\caption{Stellar velocity dispersion map for the central region of Mrk\,1066 obtained by the fit of the CO band heads around 2.3\,$\mu$m 
% using the penalized Pixel-Fitting (pPXF) method of \citet{cappellari04}. See \citet{paper2} for a discussion 
%on the stellar kinematics.} 
%\caption{Mass-weighted contribution of each SP.} 
%\label{sig}
%\end{figure}

\subsection{Observations and data reduction}\label{obs}
Mrk\,1066 was observed with the Gemini Near-Infrared Integral-Field Spectrograph \citep[NIFS;][]{mcgregor03} operating with Gemini North Adaptive Optics system ALTAIR in September 2008 under the programme GN-2008B-Q-30. Two set of observations were obtained, one at the J band, covering a spectral region from 1.15\,$\mu$m to 1.36\,$\mu$m with a two-pixel spectral resolving power  
of  6040; and another at the \kl\ band at resolving power of 5290 and covering the spectral range  2.1--2.5$\,\mu$m.  

The data reduction, described in \citet{paper1}, was performed using the {\sc gemini iraf} package and followed  standard procedures. 
The resulting calibrated data cube contains 784 spectra covering a angular region of 2\farcs8$\times$2\farcs8 at a sampling of  0\farcs1$\times$0\farcs1, which corresponds to 650$\times$650\,pc$^2$ and 23.5$\times$23.5\,pc$^2$, respectively. The angular resolution is $\approx$0\farcs15, corresponding to 35\,pc at the galaxy. 
%For more details about the observation and data reduction procedure of Mrk\,1066  see \citet{paper1}

\subsection{Analysis}\label{ana}

% We constructed our spectral base set using the  Evolutionary Population Synthesis (EPS) models of \citet{maraston05}, which include the effects of the 
% TP-AGB stars (important in the near-IR spectral region). Following \citet{rogerio09}, we chosen single stellar populations (SSPs) synthetic spectra
% covering 14 ages ($t=$0.001,0.005,0.01,0.03, 0.05, 0.1, 0.2, 0.5, 0.7, 1, 2, 5, 9 and 13~Gyr)  and four metallicities ($Z=$0.02, 0.5, 1, 2~Z$_\odot$). In 
%order to account for the continuum emission due to hot dust, we included in our spectral base eight Planck distributions with temperatures ranging from 
%700 to 1400~K, in steps of 100~K.
%% The upper limit $T=1400$~K was chosen because it is close to the sublimation temperature 
%%of the graphite, while the lower limit $T=700$~K is due to the fact that it is very hard to detect temperatures lower than this value in the near-IR 
%%\citep{rogerio09,barvainis87}. 
%We also take into account a possible effect of the featureless continuum (FC) of the AGN including a power-law of the form 
%$F_\nu\propto\nu^{-1.5}$ inthe spectral base \citep[e.g.][]{rogerio09,cid04}.  For more details on the construction of the spectral base see 
%\citet{rogerio09}.

In order to model the continuum and stellar absorption features we
used the {\sc starlight} spectral synthesis code \citep{cid04,cid05,cid09,mateus06,asari07}.  
According to the authors {\sc starlight} mixes computational
techniques originally developed for semi empirical population synthesis with ingredients from evolutionary synthesis models.

In summary, the code fits an observed spectum $O_{\lambda}$ with a combination, in different proportions, of
$N_{*}$ single stellar populations (SSPs). Following \citet{rogerio09} we use the \citet{maraston05}  Evolutionary Population Synthesis (EPS)
 models as base set (described below).
This choice is due to the fact that Maraston's models include the effects of stars in the TP-AGB phase, wich is crucial to model NIR SP 
\citep[see][]{rogerio06,rogerio07,rogerio08,rogerio09}. 
Extinction is modeled by {\sc starlight} as due to foreground dust, and
parameterized by the V-band extinction $A_V$ \citep{cid04,cid05}. Note that all the components are reddened by the same ammount.
We use the CCM \citep{ccm89} extinction law.

Basically, the code solves the following equation for a model spectrum $M_{\lambda}$ \citep{cid05}:
\begin{equation}
M_{\lambda}=M_{\lambda 0}\left[\sum_{j=1}^{N_{*}}x_j\,b_{j,\lambda}\,r_{\lambda} \right] \otimes G(v_{*},\sigma_{*})
\end{equation}
were $b_{j,\lambda}\,r_{\lambda}$ is the reddened spectrum of the $j$th SSP normalized at
$\lambda_0$; $r_{\lambda}=10^{-0.4(A_{\lambda}-A_{\lambda 0})}$ is the reddening term; $M_{\lambda 0}$ is the
synthetic flux at the normalisation wavelength; $\vec{x}$ is the population vector; $\otimes$ denotes the convolution
operator and $G(v_{*},\sigma_{*})$ is the gaussian distribution used to model the line-of-sight stellar
motions, it is centred at velocity $v_{*}$  with dispersion  $\sigma_{*}$.

The final fit is carried out with a simulated annealing plus Metropolis scheme, which searches for the
minimum of the equation \citep{cid05}:

\begin{equation}
\chi^2 = \sum_{\lambda}[(O_{\lambda}-M_{\lambda})w_{\lambda}]^2
\end{equation}
where emission lines and spurious features are masked out by fixing $w_{\lambda}$=0.

%In order to model the continuum and stellar absorption features we used the {\sc starlight} spectral synthesis code \citep{cid04,cid05}. This code fits a galaxy's spectrum  by a combination of  elements from a  ``spectral base" (described bellow),  giving as output the weights of each base element in the synthetic spectrum \citep[e.g.][]{cid04,cid05,cid09}.

The spectral base was constructed with EPS models of \citet{maraston05} as in \citet{rogerio09}, and comprise single stellar populations (SSPs) synthetic spectra covering 12 ages ($t=$0.01,0.03, 0.05, 0.1, 0.3, 0.5, 0.7, 1, 2, 5, 9 and 13~Gyr) and four metallicities ($Z=$0.02, 0.5, 1, 2~Z$_\odot$). We also included blackbody functions for temperatures in the range 700-1400\,K in steps of 100\,K \citep{rogerio09} and a  power-law ($F_\nu\propto\nu^{-1.5}$) in order to account for possible contributions from dust emission and from a featureless continuum (FC), respectively, in regions close to the nucleus \citep[e.g.][]{cid04}.

% The Maraston's models have a lower spectral resolution than our spectra and thus, we degraded the observed spectra to 
% the resolution of the models in order to investigate if the resolution difference could affect the resulting SPs.
% The comparisson between the fitting of the degraded and non-degraded spectra shows that the
% contributions of each component of the spectral base are very similar for both cases. 
%Nevertheless, the emission-line profiles in the degraded spectra are broader than in the original spectra and  thus contaminate 
% continuum regions, decreasing the number of constraints of the fitting, and thus the resulting maps are noisier 
% than the ones obtained by the fitting of the original spectra. Therefore, we chose to fit the original spectra since the results
% obtained from both, degraded and original, spectra are very similar.

\section{Results} \label{results}

\begin{figure*}
 \centering
% \begin{minipage(1\linewidth}
 \includegraphics[scale=0.62]{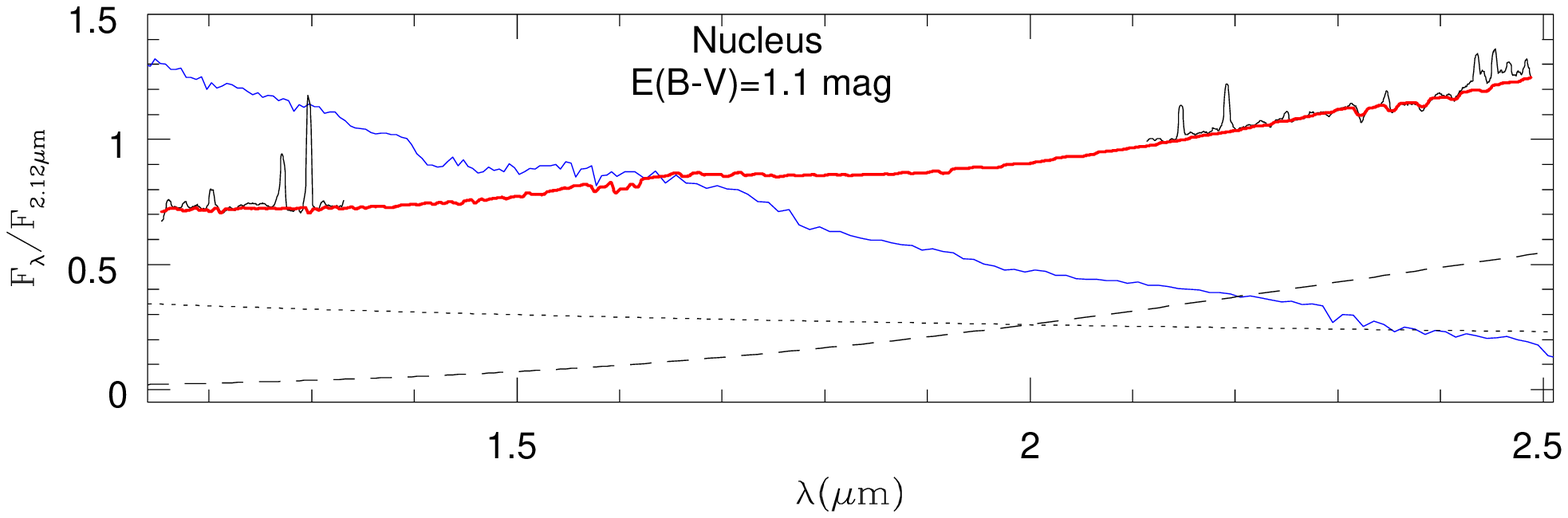}
\includegraphics[scale=0.62]{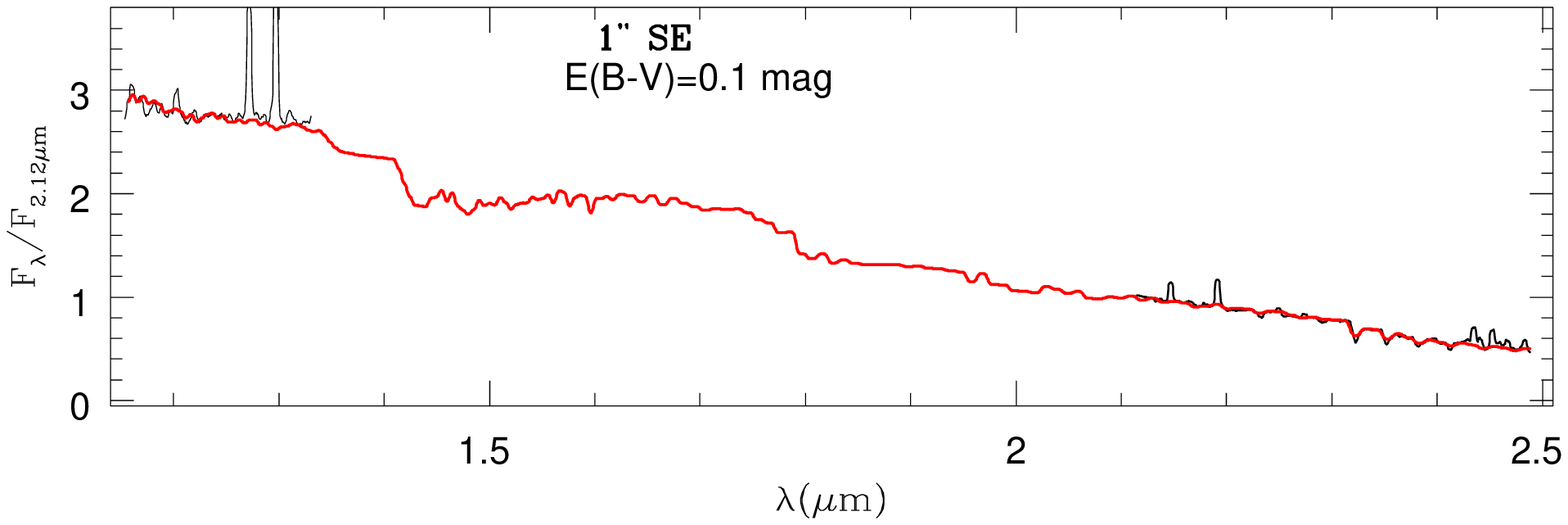}
% \end{minipage}
 \caption{Sample fits of the SPCs to the nuclear spectrum (top) and to an extra-nuclear one at 1\farcs0 south-east of the nucleus (bottom). The observed spectra are shown in black and the fits in red. In the top panel the dashed line shows the contribution of hot dust, the dotted line that of the featureless continuum and the blue continuous line shows the contribution of the combined SPCs.% Note that the individual components are not corrected by reddening.
 %The box in the top-left corner of the bottom panel shows a zoom in the CO absorption band heads after degrade the galaxy spectrum to the 
%spectral resolution of the SSPs (see text).
}
\label{sample}  
 \end{figure*}

\begin{figure*}
\centering
 \begin{minipage}{1\linewidth}
\includegraphics[scale=1.]{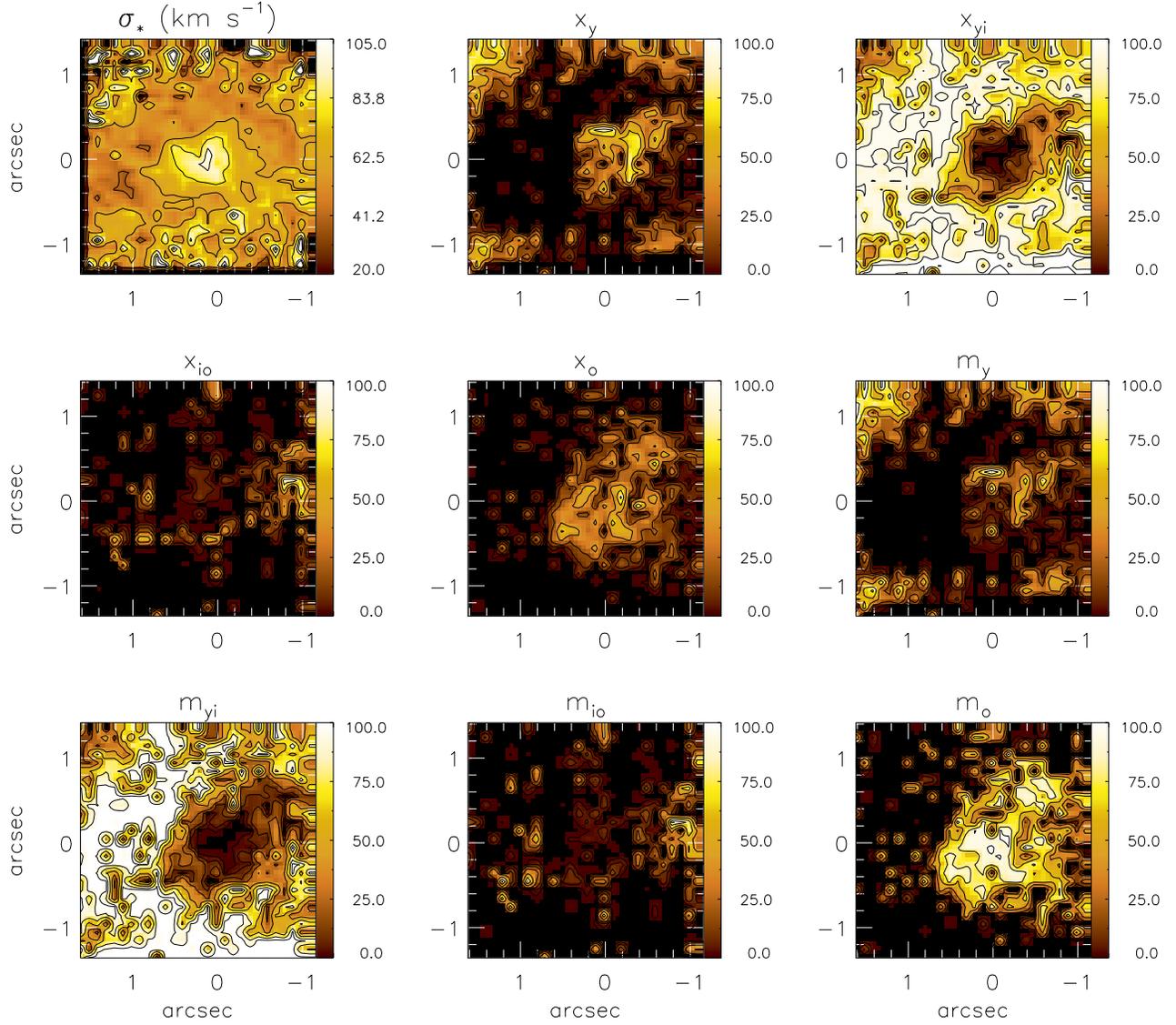}
\end{minipage}
\caption{Stellar velocity dispersion map (top-left panel) and 
spatial distributions of the percent contribution of each SPC to the flux 
at $\lambda=2.12\,\mu$m ($x_j$) and to the mass ($m_j$), where $j$ represents the age of the 
SPC: young ($y$: $\leq$ 100 Myr), young-intermediate ($yi$: 0.3--0.7~Gyr), intermediate-old ($io$: 1--2~Gyr) 
and old ($o$: 5--13~Gyr).
%
%LEFT: Spatial distributions of the percent contribution of each SPC to the flux at $\lambda=2.12\,\mu$m.  
%From top to bottom: young ( $\leq$ 100 Myr), young-intermediate (0.3--0.7~Gyr), intermediate-old (1--2~Gyr) 
%and old (5--13~Gyr) SPCs. MIDDLE: Spatial distributions of mass-weighted contribution from each SPC.
% RIGHT -- From top to bottom: Spatial distribution of the dust emission summed over all temperatures, stellar velocity dispersion map, reddening map and {\it adev} map (percent mean deviation from the spectral fit). %Black regions 
} 
\label{pop}
\end{figure*}

\begin{figure*}
\centering
 \begin{minipage}{1\linewidth}
\includegraphics[scale=1.]{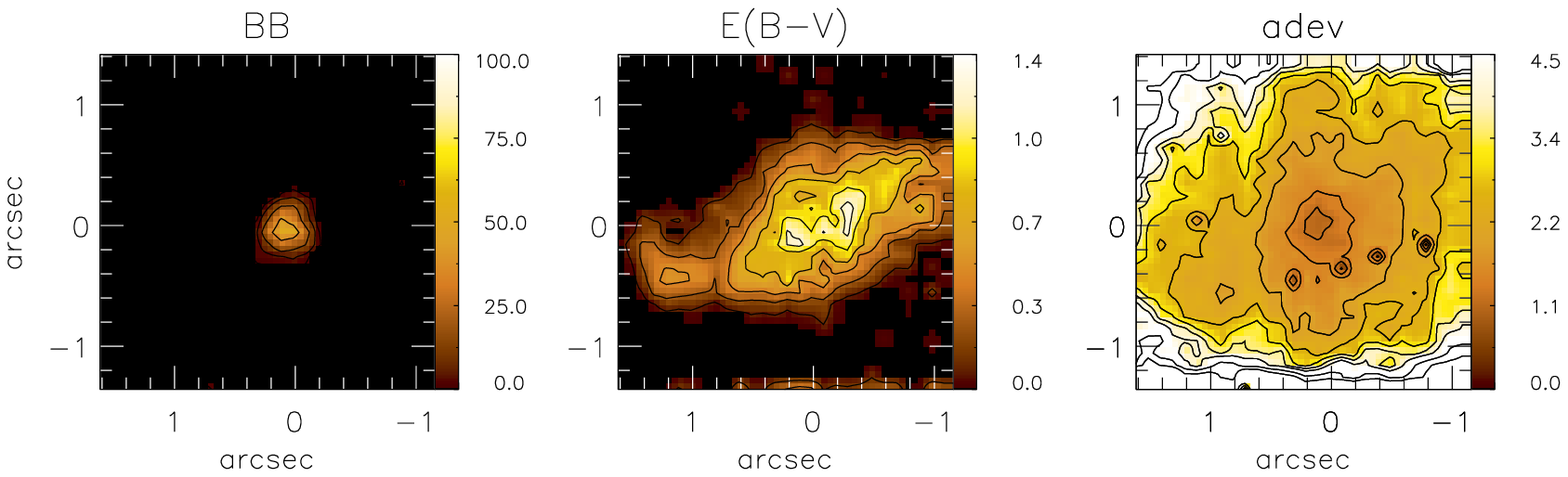}
\end{minipage}
\caption{From left to right: 
Spatial distribution of the dust emission summed over all temperatures, 
 reddening map and {\it adev} map (percent mean deviation from the spectral fit).} 
\label{bb}
\end{figure*}

%\begin{figure*}
%\centering
%\includegraphics[scale=0.65]{figs/pop_mass.ps}
%\caption{Mass-weighted contribution of each SP.}  
%\label{pop_mass}
%\end{figure*}

In Fig.~\ref{sample} we present the results of the synthesis for two spectra extracted within an aperture of 0\farcs1$\times$0\farcs1, one  at the nucleus (top panel) and the other at 1$^{\prime\prime}$ south-east of it (bottom panel). The observed (black) and synthetic (red) spectra were normalized to unit at 2.12\,$\mu$m. This wavelength was chosen because the K-band spectra  present a higher signal-to-noise ratio than those in the J-band and the spectral region near 2.12~$\mu$m is free of emission/absorption lines \citep{rogerio08}. 
%Although our observations cover only the J and K bands, we present the synthetic spectra over the whole spectral range (1.15--2.5$\mu$m), which shows that there is no 
%descontinuity between the J and K bands indicating a good flux calibration. 
In the inner 0\farcs2 (45\,pc)  radius the contribution from a FC and from dust emission are important, 
as illustrated in the top panel of Fig.~\ref{sample}, where the combined contribution of all SPCs is shown in blue, the FC as a dotted line and the composite blackbody function (sum over all temperatures) as a dashed line.
We point out that the syntetic spectra (sum of all components) was
reddened by the reddening amount detected in the observed spectra.
% We point out  that the observed and synthetic spectra have not been corrected for reddening. 
%The observed spectra is 
%shown in black in Fig.~\ref{sample}, while the resulting fit  in red. For the nucleus, where the 
%FC and the dust emission are important,  the blue line shows the contribution of the SP, the the blackbody function (sum over all temperatures) 
%is shown as a dashed line and the FC as a dotted line. Note that the individual components are not reddening corrected. 
%  As observed in this figure the resulting fit reproduces very well the continuum emission, but some stellar absorptions
% are deeper at the observed spectra than at the modeled one, although the equivalent width of the stellar absorptions 
%in both spectra are in good agreement to each other.  This may be due to the higher spectral resolution of the data than the one of the SSPs.

Following \citet{cid04}, we have binned the contribution of the SPCs ($x_j$) to the flux at 2.2$\mu$m into four age ($t$) ranges:  young ($x_y$: $t \leq 100$~Myr); young-intermediate ($x_{yi}$: $0.3 \leq t \leq 0.7$~Gyr), intermediate-old  ($x_{io}$: $1 \leq t \leq 2$~Gyr)  and old  ($x_{o}$: $5 \leq t \leq 13$~Gyr). 
%The ranges of ages were chosen following \citet{cid05} 
%-- see also \citet{rogerio09}, which binned the SPs into three groups: young, intermediate and old. Our young and old populations have the same range of ages than theirs, while the the intermediate 
%population was divided into two groups in order to better discriminate the contribution of each age. 
The percent flux contribution at 2.12\,$\mu$m from stars of  each age range are shown in 
 Fig.\,\ref{pop}:  on average, the young-intermediate SPCs dominate the continuum emission 
from the central region of Mrk\,1066, but there is significant variation over the NIFS field.
 While in regions farther than $r\sim$0\farcs7 -- 160\,pc  from the nucleus -- the contribution 
of the young-intermediate age stars reaches values of up to 100\,\%, closer to the nucleus its
 contribution is negligible. Within this region the stellar population is dominated by the old 
component, whose contribution reaches $\approx50$\,\%, followed by the young component, whose 
contribution reaches $\approx$30\,\% of the flux at 2.12\,$\mu$m. Also within this region, but
 unresolved by our observations (thus within 0\farcs15 from the nucleus) we find a contribution
 of up to 50\% from the combined blackbody components. Finally, we have found a small contribution
 from the AGN FC component, which reaches at most 15\,\% at the nucleus, also unresolved by our observations. 
%{\bf O TEXTO A SEGUIR FOI MOVIDO PARA AS DISCUSSOES E SERA REMOVIDO DAQUI: This contribution is
% nevertheless uncertain, as it is hard to distinguish a reddened young starburst 
%($t\lesssim5$\,Myr) from an AGN FC component \citep[e.g.][]{sb00,cid04,rogerio09}. 
%On the other hand, the contribution of the young population $x_y$ is dominated by stars 
%with ages of 50--100\,Myr, thus there does not seem to be a significant contribution from 
%very young stellar populations. }
The intermediate-old component contributes with less than
 20\,\% at most locations, with larger values more concentrated to the west-northwest of the nucleus.

Besides the SPC distributions, the synthesis outputs the average reddening of the stellar populations, shown in the 
central panel of Fig.\,\ref{bb}. It reaches the highest values, of up to $E(B-V)=1.4$, along the position angle 128$^\circ$,
 which is the orientation of the line of nodes of the galaxy \citep{paper2}.

The robustness of the SP fit can be measured by the percent mean deviation (\textit{adev}), $|O_\lambda-M_\lambda|/O_\lambda$, 
 where $O_\lambda$ is the observed spectrum and $M_\lambda$ is the fitted model \citep{cid04,cid05}.  
The resulting \textit{adev} map is shown in the right panel of Fig.~\ref{bb} and presents
 values $adev\lesssim2.5$\,\% at most locations, indicating that the model reproduces very well the observed spectra. 
Nevertheless, at regions close to the border of the NIFS field the $adev$  reaches values of up to 5\%. 
At these locations we found an increase in the contribution of the young component $x_y$, 
which has thus larger uncertainties than in the rest of the NIFS field and should be further investigated with better data.

\section{Discussion} \label{discussion}

In general, our results are in good agreement with those of previous near-IR studies for single aperture nuclear spectra  
\citep{rogerio09,ramosalmeida09}. The main novelty of our work is the 2D mapping of the stellar population using near-IR spectra, which shows spatial variations in the contribution of the SPCs in the inner few hundred parsecs of Mrk\,1066. 

The significance of the above variations is further enhanced by the comparison with the  $\sigma_*$ map 
presented in the top-left panel of Fig.~\ref{pop}. This map shows a partial ring 
of low $\sigma_*$ values ($\approx50\,{\rm km\,s^{-1}}$) surrounding the nuclear region (which has  $\sigma_*\,\approx100\,{\rm km\,s^{-1}}$) at $\approx 1^{\prime\prime}$ from it. Such rings are commonly observed in the central region of active galaxies \citep{barbosa06,deo06,lopes07,n4051,n7582} and  interpreted as being due to colder regions with more recent star formation than the underlying bulge. The comparison of the results from the SP synthesis of Mrk\,1066 with the $\sigma_*$ map  shows that the low $\sigma_*$-values ring is associated with the young-intermediate age SPC, while the highest  $\sigma_*$ values are associated with the old SPC, supporting the use of low stellar velocity dispersion 
as a tracer of younger stars in the bulge and confirming the interpretation of the above studies.

The flux-weighted SPC  contributions depend on the normalization point and thus the comparison with 
results from other 
spectral regions should be done with caution. A physical parameter which does not depend on 
the normalization point and spectral range used in the  synthesis is the mass of the stars. 
Thus, we constructed maps for the mass-weighted contribution of each SPC. 
The contributions for the young population ($m_y$),  young-intermediate population ($m_{yi}$), 
 intermediate-old population ($m_{io}$) 
and old population ($m_o$), are shown  in the rightmost panel of second row, 
bottom-left, bottom-central and bottom-right 
panels of Fig.~\ref{pop}, respectively. The mass-weighted contribution of the 
young population is very small over the whole field, while $m_o$ dominates within 
$\approx$160\,pc from the nucleus and $m_{yi}$ dominates in the circumnuclear ring.

The only previous 2D stellar population studies of active galaxies in the near-IR published to date are those from the group of R. I. Davies. In particular, \citet{davies07} investigated the circumnuclear star formation in nine Seyfert galaxies using near-IR IFU observations and also found circumnuclear disks of typical diameters of tens of pc  with  a ``characteristic" age in the range 10--300~Myr based on measurementes of the \br\ emission-line equivalent width,  supernova rate and mass-to-light ratio. Thus, the results we have found for Mrk\,1066 are in reasonable agreement with those found by Davies group. 
%In the case of Mrk\,1066, we obtain that stars with ages from 100 to 500~Myr dominates the flux 
%in the K band, thus in reasonable agreement with the results obtained for other Seyfert galaxies by 
%the above authors. 
Nevertheless, the methodology adopted in the present work allowed us not only to obtain a ``characteristic" age, but to also map, for the first time, the spatial distribution of stars of different ages in the central region of a Seyfert galaxy, on the basis of near-IR integral-field spectroscopy.
%As observed in Fig.~\ref{pop}, the contributions of the FC and of dust to the continuum emission at 2.12\,$\mu$m  
%peak at the nucleus and are unresolved by our observations. The FC contribution have a peak value of $\approx$20\,\% of the continuum flux,
% in good agreement with the one obtained in \citet{rogerio09}, while the  dust emission contributes 
%with up to 45\,\% of the flux at 2.12\,$\mu$m, which is simmilar to the value obtained from the fitting of the nuclear spectrum of Mrk\,1066 
%by a blackbody function \citep{paper1,ramosalmeida09}.

As presented in Sec.~\ref{ana}  our spectral base include four distinc metalicities: $Z=$0.02, 0.5, 1, 2~Z$_\odot$.
 Following \citet{cid04}, 
 we can estimate the mean flux and mass-weighted metalicity in the fit by 
\begin{equation}
 <Z_*>_L=\sum_{j=1}^{N_*} x_j Z_j
\end{equation}
and
\begin{equation}
 <Z_*>_M=\sum_{j=1}^{N_*} m_j Z_j,
\end{equation}
where $<Z_*>_L$ is the flux-weighted mean metalicity, $<Z_*>_M$ is the mass-weighted mean metalicity and $N_*$ 
is the number of SSP in the spectral base. In the case of Mrk\,1066 we obtain a nearly solar metalicity -- $<Z_*>_L=0.021$
 and $<Z_*>_M=0.017$ -- in good agreement with those found by \citet{rogerio09}
 from near-IR spectral synthesis using a single aperture nuclear spectrum.

In further support of the results of the synthesis, we found that the average reddening map
 derived for the stellar population (central panel of Fig.\,\ref{bb}) is in close agreement to the one derived for the  narrow-line region using emission-line ratios \citep{paper1}, presenting a similar S-shaped structure. 

The synthesis also confirmed the presence of an unresolved blackbody (BB) component at the nucleus,
 which we had found in \citet{paper1} from the fit of the nuclear spectrum by a BB function plus a power-law. 
On the basis of its luminosity, we have followed the calculations as in \citet{rogerio09} to estimate a total 
dust mass of $M_{\rm HD}\approx1.9\times10^{-2}$~M$_\odot$, which is in good agreement with the value obtained in 
\citet{paper1} and with those observed for other Seyfert galaxies  \citep[e.g.][]{n7582,n4151,rogerio09,ardila06,ardila05}.

 As discussed in Sec.~\ref{results}, we found that the FC component contributes with $\sim$ 15\,\% 
of the 2.12\,$\mu$m nuclear continuum, which is in good agreement with the result found by 
\citet{rogerio09}. This contribution is nevertheless uncertain,
 as it is hard to distinguish a reddened young starburst 
($t\lesssim5$\,Myr) from an AGN FC component \citep[e.g.][]{sb00,cid04,rogerio09}. 
On the other hand, the contribution of the young population $x_y$ is dominated by stars 
with ages of 50--100\,Myr, thus there does not seem to be a significant contribution from 
very young stellar populations, suggesting that the FC component originates from the 
emission of the AGN. This interpretation is supported by the detection of weak broad components 
for the Pa$\beta$ and Br$\gamma$ emission lines for the nuclear spectra of Mrk\,1066 \citep{paper1,veilleux97}. 
 However, spectro-polarimetric studies, 
show no evidence for the hidden Seyfert 1 nucleus in Mrk\,1066 \citep[e.g.][]{bian07}. Thus, 
the origin of the FC in Mrk\,1066 must be further investigated.

We finally point out one caveat of the stellar population synthesis in the near-IR. In \citet{paper1}, on the basis of emission-line ratios, 
we have identified a large star-forming region at 0\farcs5 south-east of the nucleus. In order to be active,
 this region should have an age $\lesssim$10~Myr
 \citep{n7582,dors08,diaz07,kennicutt89}. Nevertheless, our synthesis did not find a young stellar component 
at this location, indicating that spectral synthesis in the near-IR (at least in the J and K bands) is not a 
good tracer of very recent  star formation (age $\le$\,10\,Myr). 
This result is expected since the  young (blue) population has its peak emission in ultraviolet/optical
 wavelengths, while in the near-IR its contribution to the flux is 
very small, being hard to detect this component in this wavelength region.

 In order to have a complete census of the stellar population components it is thus essential to study the stellar 
content of the central region of active galaxies using distinct methods and spectral regions.
% \citep[e.g.][]{n7582,rogerio09,rogerio08,dors08,davies07}. {\it Estas referencias nao estao adequadas aqui, usar referencias com sintese no otico e UV!}

%In order to better compare the contribution of each stellar component, bu only along one direction, we present one-dimensional cuts for the flux- and mass-weighted 
%contributions of each stellar population along the position angle 128$^\circ$, which is the orientation of the major axis of the galaxy \citep{paper2}. These cuts 
%are presented in Fig.~\ref{cut} and were obtained by averaging the contribution of each component within a pseudo-slit with 0\farcs3 width centred at the nucleus. 
%This figure clearly shows that the old population dominates in the mass-weighted cuts (bottom panel), although the contribution of the intermediate-age population is higher 
%than 30\,\% in most regions. In the flux-weighted cuts (top panel), the intermediate-age population presents the higher contribution, but the old population 
%also contributes significantly. 

%\begin{figure}
%\centering
%\includegraphics[scale=0.5]{figs/cut.ps}
%\caption{One-dimensional cut for the stellar population contributions along the PA=128$^\circ$ within a pesudo-slit with 0\farcs3 whidth for flux- (top)
% and mass-weighted contributions. Continuous lines: old; dotted lines:  intermediate-old; dashed lines: young-intermediate 
%and dot-dashed lines: young.}  
%\label{cut}
%\end{figure}

\section{Conclusions} \label{disc}

% Near-IR spectral synthesis of the nuclear spectrum of Mrk\,1066 for an aperture of 0\farcs8 radius shows that the old population represent $\sim$70\,\%, the intermediate population contributes with almost 30\% and the young population with 
%less than 1\,\% of the mass of the stars \citep{rogerio09}. \citet{ramosalmeida09}  found that the average age of the circumnuclear SPs of a sample 
%of five Seyfert galaxies, including Mrk\,1066, is 
%in the range 0.1--1~Gyr, based on near-IR spectra for an 1\farcs5 aperture. Their results are similar to the ones obtained in \citet{rogerio09} for the flux-weighted 
%SPs.  Extracting the contributions of each population in the same apertures than the ones used in these works we found that they are very similar each to other.
%In optical wavelengths the young SP dominates the nuclear continuum emission, although the intermediate 
%and old populations also present significant contribution to the optical light  \citep{gd01,raimann03}. Thus, the comparison of the near-IR with the optical results shows 
%that intermediate and old stars dominate the near-IR light, while at optical wavelengths the young SP dominates, what is expected since a small fraction 
%of AGB/TP-AGB stars contribute significantly to the near-IR light \citep{maraston05}.

%\section{Final Remarks}

The present work reports for the first time  spectral synthesis in the near-IR with 2D coverage for the nuclear region of a Seyfert galaxy (Mrk\,1066) within the inner $\approx$\,300\,pc at a spatial resolution of $\approx$\,35\,pc. We have mapped the distribution of stellar population components of different ages and of their average reddening. The main conclusions of this work are:

\begin{itemize}
\item 
The age of the dominant stellar population presents spatial variations: 
the flux and mass contributions within the inner $\approx$\,160~pc are dominated by old stars 
($t \ge$5\,Gyr), while intermediate age stars ($0.3 \leq t \leq 0.7$~Gyr) dominate in the circumnuclear region;

\item 
There is a spatial correlation between the distribution of the intermediate
age component and  low stellar velocity dispersion values
which delineate a partial ring around the nucleus. Similar structures have been
found around other active nuclei, and our result for Mrk\,1066 suggests that
these nuclear rings (and in some cases disks) are formed by intermediate age stars.

\item 
There is an unresolved dusty structure at the nucleus with mass
 $M_{\rm HD}\approx1.9\times10^{-2}$~M$_\odot$, which may be the hottest part of the dusty torus postulated by the unified model of AGN 
and a small contribution from a power-law continuum ($\approx$15\,\% of the flux at 2.12\,$\mu$m);

\item 
The near-IR synthesis seems not to be sensitive to very recent star formation (with $t\lesssim5\,$Myr),  
reinforcing the importance of multi-wavelength stellar population studies of the central region  of active galaxies.

\end{itemize}

\acknowledgments 
Based on observations obtained at the Gemini Observatory, 
which is operated by the Association of Universities for Research in Astronomy, Inc., under a cooperative agreement with the 
NSF on behalf of the Gemini partnership: the National Science Foundation (United States), the Science and Technology 
Facilities Council (United Kingdom), the National Research Council (Canada), CONICYT (Chile), the Australian Research 
Council (Australia), Minist\'erio da Ci\^encia e Tecnologia (Brazil) and south-eastCYT (Argentina).  
This work has been partially supported by the Brazilian institutions CNPq and CAPES.

{}   
\clearpage
\end{document}